\documentclass[journal]{IEEEtran}
\renewcommand{\baselinestretch}{1.0}
\renewcommand{\theenumii}{\theenumi.\arabic{enumii}}

\usepackage[utf8]{inputenc}
\usepackage{subfigure}
\usepackage{amssymb}
\usepackage{color,soul}
\usepackage[T1]{fontenc}
\usepackage{multicol}
\usepackage{multirow}
\usepackage{textcomp}
\usepackage{multirow}
\usepackage{bigdelim}
\usepackage{graphicx}
\usepackage{epstopdf}
\usepackage{amssymb}
\usepackage{amsmath}
\usepackage{psfrag}
\usepackage{amsthm}
\usepackage{subfigure}
\usepackage{algorithmic,algorithm}
\usepackage{mathrsfs}
\usepackage{framed}
\usepackage{color}
\usepackage{multirow,enumerate}
\usepackage{bigdelim}
\definecolor{shadecolor}{rgb}{0.92,0.92,0.92}
\usepackage{soul}
\usepackage{color}
\usepackage{cuted}
\usepackage{cite}
\usepackage{diagbox}
\theoremstyle{definition}

\newtheorem{theorem}{Theorem}

\newtheorem{definition}{Definition}

\newtheorem{lemma}{Lemma}

\newtheorem{remark}{Remark}

\makeatletter
\newcommand{\vast}{\bBigg@{3.2}}
\newcommand{\Vast}{\bBigg@{4.5}}

\makeatother

\graphicspath{{Figure/}}

\begin{document}

\title{\LARGE \bf Data-Driven Predictive Control Towards Multi-Agent Motion Planning With Non-Parametric Closed-Loop Behavior Learning }

\author{
	Jun Ma, 
	Zilong Cheng, 
	Wenxin Wang, 
	Abdullah Al Mamun, \\
	Clarence W. de Silva,  \IEEEmembership{Life Fellow,~IEEE,}
	and Tong Heng Lee	
	\thanks{Jun Ma is with the Robotics and Autonomous Systems Thrust and the Department of Electronic and Computer Engineering, The Hong Kong University of Science and Technology, China. Email: \tt\small jun.ma@ust.hk.}
	\thanks{Zilong Cheng, Wenxin Wang, Abdullah Al Mamun, and Tong Heng Lee are with the Department of Electrical and Computer Engineering, National University of Singapore, Singapore 117583. Email: \tt\small zilongcheng@u.nus.edu; wenxin.wang@u.nus.edu; eleaam@nus.edu.sg; eleleeth@nus.edu.sg.}
	\thanks{Clarence W. de Silva is with the Department of Mechanical Engineering, University
		of British Columbia, Vancouver, BC, Canada V6T 1Z4. Email: \tt\small desilva@mech.ubc.ca.}
	\thanks{This work has been submitted to the IEEE for possible publication. Copyright may be transferred without notice, after which this version may no longer be accessible.}
}

\markboth{}
{J. Ma\MakeLowercase{\textit{et al.}}}
\maketitle

\begin{abstract}

	\textcolor{black}{In many specific scenarios, accurate and effective system identification is a commonly encountered challenge in the model predictive control (MPC) formulation. As a consequence, the overall system performance could be significantly weakened in outcome when the traditional MPC algorithm is adopted under those circumstances when such accuracy is lacking. This paper investigates a non-parametric closed-loop behavior learning method for multi-agent motion planning, which underpins a data-driven predictive control framework. Utilizing an innovative methodology with closed-loop input/output measurements of the unknown system, the behavior of the system is learned based on the collected dataset, and thus the constructed non-parametric predictive model can be used to determine the optimal control actions. This non-parametric predictive control framework alleviates the heavy computational burden commonly encountered in the optimization procedures typically in alternate methodologies requiring open-loop input/output measurement data collection and parametric system identification. 
		The proposed data-driven approach is also shown to preserve good robustness properties. 
		Finally, a multi-UAV system is used to demonstrate the highly effective outcome of this promising development.}

\end{abstract}


\section{Introduction}

Highly efficient and effective motion planning 
has certainly been the subject of significant and substantial efforts in the research literature, attracting increasing attention from researchers and engineers. 
In particular, due to the fact that autonomous systems are generally under various constraints, optimization-based methods have  been actively developed. As one of the competing optimization-based methods, the model predictive control (MPC) methodology has been widely used to tackle such resulting constrained sequential planning problems~\cite{mayne2000constrained,karg2020efficient}. It is noteworthy that the MPC approach addresses the trajectory generation problem in a highly effective manner, and system requirements can be explicitly expressed as equality and inequality constraints as part of a control synthesis problem. Generally, most of these MPC-based works focus on the motion tasks with the system model obtained from system identification processes. However, in certain scenarios where the system identification is rather costly or the parameters cannot be identified accurately~\cite{kozlowski2012modelling}, the system performance could be diminished. 
Considering this issue, there have been several rather promising preliminary studies on data-driven approaches~\cite{jiang2018data,mu2016data,zhang2018data}. Essentially, these data-driven approaches avoid the necessity for very precise system identification; and thus they can rather effectively accommodate the existence of parametric uncertainties~\cite{ma2019data}. A series of representative non-parametric predictive control approaches are reported in~\cite{annual}. Particularly, it is remarkable that the most recent advances involve the development of a so-called data-enabled predictive control (DeePC) algorithm~\cite{coulson2019data,coulson2019regularized,coulson2021distributionally}, where input/output measurement data are collected by drawing the input sequence from an arbitrary input sequence as long as it is persistently exciting. However, a direct and adverse consequence that possibly incurs is the extremely large output signals that are caused in instances where the open-loop is an unstable system; and this indeed further aggravates the computational burden and challenges to derive a reliable solution. 
These challenges notwithstanding, it is noteworthy that (along similar lines) a data-driven model-free adaptive predictive control method has also been shown in~\cite{hou2016lazy}, where it can be employed in a class of discrete-time single-input and single-output nonlinear systems. 
Furthermore, another piece of interesting work in~\cite{berberich2020data} introduces a robust data-driven model predictive control for linear time-invariant (LTI) systems, and the approach suitably assures the exponential stability of the closed-loop system with respect to the noise level. Meanwhile, tremendous efforts have likewise been dedicated to reinforcement learning, which is considered as a paradigm that trains an agent to take optimal actions (as measured by the total cumulative reward achieved) in an environment through interactions~\cite{he2020reinforcement}. 
Nevertheless, reinforcement learning is known to be prone to suffer from fatal failure (because trial-and-error is required), and also a large number of data samples are required for training processes.

\textcolor{black}{This paper investigates a non-parametric closed-loop behavior learning method for multi-agent motion planning under a data-driven predictive control framework. 
	With an innovative modified methodology with closed-loop input/output measurements 
	that comply with the appropriate persistency of excitation condition, a non-parametric closed-loop behavior learning routine is presented for motion planning of multi-agent systems under a data-driven architecture, and this relieves the computational burden encountered because of the closed-loop input/output measurements. 	
	In comparison, alternate methodologies 
	requiring open-loop input/output measurement data collection and parametric system identification 
	are typically more cumbersome and unwieldy, 
	because extremely large signals in the target system could be encountered arising from situations involving the injection of an input signal to, say, an open-loop unstable system.
	Also, compared with the model-based counterpart, it circumvents the necessity of precise system modeling. Thus, it can effectively accommodate the inevitable presence of parametric uncertainties (arising from various practical uncertainties and imperfections) such that iterative improvements in system performance are possible. }



\section{Problem Statement}

For a multi-agent motion planning problem, if precise system models of all the agents are available, it is highly attractive to use the MPC to realize the control objective. For all $\tau=t,t+1,\dotsm,t+T-1$, the LTI system dynamics can be expressed as
\begin{IEEEeqnarray}{rCl}~\label{eq:dynamics}
	x(\tau+1)&=&Ax(\tau)+Bu(\tau)\nonumber\\
	y(\tau) &=& Cx(\tau)+Du(\tau),
\end{IEEEeqnarray}
where $A\in\mathbb R^{n\times n}, B\in\mathbb R^{n\times m}, C\in\mathbb R^{q\times n}, D\in\mathbb R^{q\times m}$ are the state matrix, input matrix, output matrix, and feedthrough matrix, respectively; $x(\tau)\in\mathbb R^n,u(\tau)\in\mathbb R^{m}, y(\tau)\in\mathbb R^q$ are the state vector, input vector, and output vector, respectively; $t$ is the initial time; $T$ is the prediction horizon. 

If the weighting parameters in terms of the state variables and control input variables are chosen to be time-invariant, and a tracking problem is considered in the objective function, the MPC problem in terms of each agent with a given quadratic objective function can be formulated as
\begin{IEEEeqnarray*}{l}\label{eq:opt_individual_1}
	\min \quad
	\displaystyle\sum_{\tau =t}^{t+T} \Big(\big(x(\tau)-r(\tau)\big)^T\hat Q\big(x(\tau)-r(\tau)\big)+u(\tau)^T\hat Ru(\tau)\Big)\\
	\operatorname{subject\ to} \quad x(\tau+1)=Ax(\tau)+Bu(\tau)\\
	\,\,\,\,\quad\quad\quad\quad\quad\tau=t,t+1,\dotsm,t+T-1, \IEEEyesnumber
\end{IEEEeqnarray*}
where $\hat Q\in\mathbb S^{n}_+$ and  $\hat R\in\mathbb S^{m}_{+}$ are the weighting matrices for the state variables and control input variables, respectively; $r(\tau)$ is the reference signal at the time $\tau$.

To denote the optimization problem in a compact form, we define the optimization variables $x$ and $u$ in terms of the state variables and control input variables, respectively, where
\begin{IEEEeqnarray*}{rCl}
	x&=&\big(x(t+1),x(t+2),\dotsm,x(t+T)\big)\in \mathbb R^{nT}\\
	u&=&\big(u(t),u(t+1),\dotsm,u(t+T-1)\big)\in\mathbb R^{mT}. \yesnumber
\end{IEEEeqnarray*}
Next, define the reference vector $r$ and the weighting matrices $Q,R$ with respect to the whole prediction horizon as
\begin{IEEEeqnarray*}{rCl}
	r&=&\big(r(t+1),r(t+2),\dotsm,r(t+T)\big)\in\mathbb R^{nT}\\
	Q&=&\operatorname{diag}\Big(\underbrace{\hat Q,\hat Q,\dotsm,\hat Q}_{T}\Big)\in\mathbb R^{nT\times nT}\\
	R&=&\operatorname{diag}\Big(\underbrace{\hat R,\hat R,\dotsm,\hat R}_{T}\Big)\in\mathbb R^{mT\times mT}.\yesnumber
\end{IEEEeqnarray*}
Then the optimization problem is equivalent to
\begin{IEEEeqnarray*}{rl}
	\min \quad
	& (x-r)^TQ(x-r)+u^TRu\\
	\text{subject to} \quad &
	x=Gx^t+Hu,\yesnumber
\end{IEEEeqnarray*}
where $x^t$ is the initial state variables of the system, $G$ and $H$ are the matrices for system dynamic constraints. 

To ensure the collision avoidance constraints and also consider the box constraints on the state variables and control input variables, we generalize the single-agent tracking problem to a multi-agent scenario with $N$ agents, which yields
\begin{IEEEeqnarray*}{rl}~\label{eq:revi}
	\min \quad
	& \displaystyle \sum_{i=1}^N \Big((x_i-r_i)^TQ_i(x_i-r_i)+u^T_iR_iu_i\Big)\\
	\text{subject to} \quad
	&x_i=G_ix_{i}^t+H_iu_i\\
	&\|M  x_i-M x_j \|_2 \leq d_\textup{safe} \\
	&x_i \in\mathcal X_i \\&u_i \in \mathcal U_i\\
	&\forall i=1,2,\dotsm,N, \\&\forall j=2,3\dotsm,N, j>i,\yesnumber
\end{IEEEeqnarray*} 
In~\eqref{eq:revi}, $x_i$ represents the state vector of each agent, and the position is one of the state variables in $x_i$ as typically formulated in the tracking problem. $d_{\text{safe}}$ denotes the safe distance among the agents and $M$ is a matrix for the purpose of extracting the position state variables from the state vector. $\mathcal X_i$ and $\mathcal U_i$ denote the box constraints with respect to the input vector $u_i$ and state vector $x_i$. Note that the agents can be either homogeneous sharing the same dynamics or heterogeneous sharing the different dynamics in~\eqref{eq:revi}, as represented by $G_i$ and $H_i$.


\section{Non-Parametric Closed-Loop Behavior Learning for Multi-Agent Motion Planning}

\subsection{Input/Output Measurement Data Collection Under Closed-Loop Control}


As follows, the pertinent definition of a sequence that is stated as being persistently exciting is given. 
\begin{definition}
	A sequence $u=\{u_k\}_1^T$ with $u_k \in \mathbb{R}^m$ is persistently exciting of order $L$ if the generalized Hankel matrix
	\begin{IEEEeqnarray}{rCl}
		\mathscr{H}_L(u)=
		\begin{bmatrix}
			u_1 & \cdots & u_{T-L+1} \\ 
			\vdots & \ddots & \vdots \\
			u_L & \cdots & u_{T}
		\end{bmatrix}
	\end{IEEEeqnarray} has full row rank.
\end{definition}

\begin{remark}
	In general, a Hankel matrix is a square matrix, where each ascending skew-diagonal from left to right is constant. In this work, the term generalized Hankel matrix is used because it is not necessary to be a square matrix.
\end{remark}

In the remaining text, $T_{p}$ and $T_{f}$ denote the length of past data and future data, respectively. $T_\textup{num}$ denotes the number of input/output measurements, $L$ represents the exciting order. Recall that $n$, $m$, and $q$ represent the number of the state variables, input variables, and output variables, respectively.

To generate the input/output measurement data for each agent, we inject a signal to the input channels of each agent. In this work, we consider the system input signal as the noise generated from the uniform distribution, and the input signal is denoted by $u_d$. At the same time, the corresponding system output signal is measured, which is denoted by $y_d$. 
Referring to~\cite{coulson2019data}, we partition the collected data into the past data and the future data. Here, we define
\begin{IEEEeqnarray*}{rCl}
	U=\begin{bmatrix}
		U_p\\U_f 
	\end{bmatrix}&= \mathscr{H}_{T_{p}+T_{f}} (u_d), \yesnumber\\
	Y=\begin{bmatrix}
		Y_p\\Y_f 
	\end{bmatrix}&= \mathscr{H}_{T_{p}+T_{f}} (y_d),  \yesnumber
\end{IEEEeqnarray*}
where $U_p$ and $Y_p$ comprise the first $T_{p}$ block rows of the corresponding generalized Hankel matrix, $U_f$ and $Y_f$ comprise the last $T_{f}$ block rows of the corresponding generalized Hankel matrix; subscripts $(\cdot)_p$ and $(\cdot)_f$ denote the past data and the future data, respectively.
It is pertinent to note that matrices $U$ and $Y$ are collected offline, and more specificly, they are organized as follows:
\begin{IEEEeqnarray}{rCl}
	U&=\begin{bmatrix}
		u_1^d &\cdots& u_{T_\textup{num}-T_{f}-T_{p}+1}^d \\
		\vdots & \ddots & \vdots\\
		u_{T_{p}}^d &\cdots & u_{T_\textup{num}-T_{f}}^d\\
		u_{T_{p}+1}^d &\cdots & u_{T_\textup{num}-T_{f}+1}^d\\
		\vdots & \ddots & \vdots\\
		u_{T_{p}+T_{f}}^d &\cdots & u_{T_\textup{num}}^d\\
	\end{bmatrix} , \\
	Y&=\begin{bmatrix}
		y_1^d &\cdots& y_{T_\textup{num}-T_{f}-T_{p}+1}^d \\
		\vdots & \ddots & \vdots\\
		y_{T_{p}}^d &\cdots & y_{T_\textup{num}-T_{f}}^d\\
		y_{T_{p}+1}^d &\cdots & y_{T_\textup{num}-T_{f}+1}^d\\
		\vdots & \ddots & \vdots\\
		y_{T_{p}+T_{f}}^d &\cdots & y_{T_\textup{num}}^d\\
	\end{bmatrix}.
\end{IEEEeqnarray}
Notably, we have $U\in\mathbb{R}^{ m(T_{p}+T_{f})\times (T_\textup{num}-T_{p}-T_{f}+1)}$ and $Y \in\mathbb{R}^{q(T_{p}+T_{f})\times (T_\textup{num}-T_{p}-T_{f}+1)}$.

As noted in~\cite{coulson2019data}, to ensure the signal is persistently exciting of order $L$, the minimum number of input/output measurement data is $T_\textup{min}=(m+1)L-1$. Furthermore, to guarantee that the column span of the generalized Hankel matrix with respect to the input/output measurement data is the whole behavior space in the behavioral view, we have $L=T_{p}+T_{f}+n$.

Typically, injecting a random signal to an unstable open-loop system leads to a significantly large output signal, which unpleasantly increases the condition number of the generalized Hankel matrix. Essentially, a large condition number can dramatically slow down the convergence of the optimization process in the following numerical procedures, and cause inexact solutions due to the existence of numerical errors. Thus, in this paper, we propose the use of closed-loop input/output measurement data to overcome this shortcoming, where a prescribed controller is designed to make sure the closed-loop system is stabilizing. 

Assume there is a controller $K$ stabilizing the closed-loop system, the system input during the data measurement process can be denoted by $u_d=Ky_d+u_r$, where $u_r$ is a random system input vector to ensure the full row rank of the corresponding generalized Hankel matrix. To facilitate the development of the data-driven approach, the following theorem is introduced.
\begin{theorem}~\label{lemma4}
	Assume the input sequence is persistently exciting and there exists a controller $K$ stabilizing the closed-loop system (denoted by ${G}_c$), it follows that
	\begin{IEEEeqnarray*}{rCl}
		y_d &=& G_c u_r\\
		u_d &=& Ky_d+u_r,\yesnumber
	\end{IEEEeqnarray*}
	where $u_r$ is a random system input vector to ensure the full row rank of the corresponding generalized Hankel matrix.
	Then the generalized Hankel matrix with respect to the measurement data $(u_d,y_d)$ spans the whole behavior space.
\end{theorem}
\noindent{\textbf{Proof of Theorem~\ref{lemma4}.}} 
To prove Theorem~\ref{lemma4}, the following lemma is introduced, which serves as the basis to derive many typical data-driven predictive control methodologies.

\begin{lemma}~\cite{willems2005note}\label{lemma1.1}
	Consider a controllable system $\mathcal{B}\in \mathcal{L}^w$. Let $\tilde u: [1,T]\rightarrow\mathbb{R}^{\textup{m}(\mathcal{B})}$, $\tilde y: [1,T]\rightarrow\mathbb{R}^{\textup{q}(\mathcal{B})}$, and $\tilde w=\begin{bmatrix}
		\tilde u^T & \tilde y^T
	\end{bmatrix}^T.$
	Assume that $\tilde w \in \mathcal{B}|_{[1,T]}$. Then, if $\tilde u$ is persistently exciting of order $L+\textup{n}(\mathcal{B})$, it yields that $\textup{columnspan}(\mathscr{H}_L(\tilde w))= \mathcal{B}|_{[1,L]}$, where $\textup{n}(\cdot)$, $\textup{m}(\cdot)$, and $\textup{q}(\cdot)$ denote the state cardinality, input cardinality, and output cardinality of a system, respectively.
\end{lemma}

From Lemma~\ref{lemma1.1}, under the persistency of excitation assumption for the input sequence, the whole behavior space can be expressed as the product space of two sub-behavior spaces in terms of the input signals and output signals, respectively. Furthermore, it means that all trajectories can be constructed from a finite number of past trajectories. Then it is straightforward to prove Theorem~\ref{lemma4}.  \hfill{\qed}


\begin{remark}
	\textcolor{black}{Under the stable condition, different values of $K$ have the same effect on the accuracy of the prediction model and the stability of predictive control, as long as the data is under the condition of persistent excitation.}
\end{remark}

\begin{remark} 
	The data included in the generalized Hankel matrix with respect to the original open-loop system is from the measurement of $u_d$ instead of $u_r$. Since $u_r$ is drawn as a random vector from a distribution, the columns of the generalized Hankel matrix are linearly independent.
\end{remark}

\subsection{Data-Driven Predictive Control Algorithm for Multi-Agent Motion Planning}
With the collected data, we can learn the behavior of the system and build a non-parametric predictive model for the determination of optimal control actions. The following theorem generalizes the results in \cite{coulson2019data} to a multi-agent scenario using alternate closed-loop measurement data.
\begin{theorem}~\label{thm:new}
	Let the input of the original open-loop system $u_{d,i}=K_iy_{d,i}+u_{r,i}$ and partition $u_{d,i}=[u_{p,i}^T\,\,\, u_{f,i}^T]^T$, $y_{d,i}=[y_{p,i}^T\,\,\, y_{f,i}^T]^T$. Then, with $W_i = [u_{p,i}^T\,\,\, y_{p,i}^T\,\,\,u_{f,i}^T\,\,\,y_{f,i}^T]^T$, and under the persistency of excitation assumption, 
	the following input/output relationship can be established:
	\begin{IEEEeqnarray}{rCl}\label{eq:con_data}
			W_i g_i =\begin{bmatrix}
				u_{p,i}^T & 	y_{p,i}^T & u_i^T& y_i^T
			\end{bmatrix}^T,
		\end{IEEEeqnarray}
		where $u_{p,i}$ and $y_{p,i}$ denote the most recent input signal and output signal of the $i$-th agent, respectively, $u_i$ represents the optimal control input sequence and $y_i$ is the corresponding output sequence. Then there exists a decision variable $g_i\in\mathbb{R}^{ T_\textup{num}-T_{p}-T_{f}+1}$ such that~\eqref{eq:con_data} holds, under the condition that no measurement noise exists in the input/output closed-loop measurement data.
	\end{theorem}
	
	\noindent{\textbf{Proof of Theorem~\ref{thm:new}.}} 
	First, the following lemma is introduced to complete the proof.
	\begin{lemma}~\cite{berberich2020trajectory}\label{lemma2}
		Suppose $\big(u^d,y^d\big) = \{u_k^d,y_k^d\}_{1}^{T}$ is a trajectory of an LTI system $G_o$, where $u^d$ is persistently exciting of order $L+n$. Then, $(\bar u, \bar y) = \{\bar u_k, \bar y_k\}_{1}^{L}$ is a trajectory of $G_o$ if and only if there exists $\alpha\in \mathbb{R}^{T-L+1}$ such that  
		\begin{IEEEeqnarray}{rCl}
			\begin{bmatrix}
				\mathscr{H}_L(u^d)\\ \mathscr{H}_L(y^d)
			\end{bmatrix}\alpha = 	\begin{bmatrix}
				\bar u \\ \bar y
			\end{bmatrix}.
		\end{IEEEeqnarray}
	\end{lemma}
	
	Under the condition that no measurement noise exists in the input/output closed-loop measurement data, it is straightforward that the column span of the generalized Hankel matrix with respect to the input/output measurement data is the whole behavior space, and then it follows from Lemma~\ref{lemma2} that the vector $g$ always exists. Specifically, pertinent details on the proof of input/output relationship can be found in~\cite{coulson2019data}. 
	\hfill{\qed}

	Notice that the results presented above are in terms of data collection and input/output relationship construction for each single agent.
	Consequently, for the multi-agent motion planning problem investigated in this work, the optimization problem is formulated as
	\begin{IEEEeqnarray*}{rl}~\label{eqn:fin}
		\min \quad
		& \displaystyle \sum_{i=1}^N \Big((y_i-r_i)^TQ_i(y_i-r_i)+u^T_iR_iu_i\Big)\\
		\operatorname{subject\ to} \quad
		&W_i g_i =\begin{bmatrix}
			u_{p,i}^T & 	y_{p,i}^T & u_i^T& y_i^T
		\end{bmatrix}^T  \\
		& \|y_i-y_j \|_2 \leq d_\textup{safe}\\
		&y_i \in\mathcal Y_i \\&u_i \in \mathcal U_i\\
		&\forall i=1,2,\dotsm,N,  \\
		&\forall j=2,3,\dotsm,N,j>i,\yesnumber
	\end{IEEEeqnarray*}
	where the feasible set $\mathcal Y_i$ denotes the box constraints on the system output. To summarize the above descriptions and discussions, Algorithm~1 is given.

	
	
	\begin{algorithm}[t]
		\label{algo}
		\caption{Proposed Algorithm for Multi-Agent Motion Planning}
		\begin{algorithmic}[1]\label{algorithm}
			\REQUIRE
			$Q_i$, $R_i$, $d_{\text{safe}}$, $T_{\text{num},i}$, $u_{p0,i}$, $y_{p0,i}$.
			\STATE \textbf{Step 1:} Design a controller $K_i$ that stabilizes the closed-loop system for each agent $i$.
			\STATE \textbf{Step 2:} Construct the matrix $W_i$.
			\FOR {$k=1,2,\dotsm,T_{\text{num},i}-T_{p,i}-T_{f,i}+1$}
			\STATE Inject random signal $u_{r,i}$ into the closed-loop system under controller $K$.
			\STATE Measure the system output $y_{d,i}$.
			\STATE Compute the input of the original open-loop system $u_{d,i}=K_iy_{d,i}+u_{r,i}$ and partition $u_{d,i}=[u_{p,i}^T\,\,\, u_{f,i}^T]^T$, $y_{d,i}=[y_{p,i}^T\,\,\, y_{f,i}^T]^T$.
			\STATE Construct the matrix $W_i = [u_{p,i}^T\,\,\, y_{p,i}^T\,\,\,u_{f,i}^T\,\,\,y_{f,i}^T]^T$.
			\ENDFOR
			\STATE \textbf{Step 3:} Implement the iterative predictive control.
			\FOR {$\tau = t,t+1,\dotsm,t+T$}
			\IF {$\tau == t$}
			\STATE Set $u_{p,i}=u_{p0,i}$ and $y_{p,i}=y_{p0,i}$.
			\ELSE
			\STATE Set $u_{p,i}$ to be the most recent system input and $y_{p,i}$ to be the most recent system output.
			\ENDIF
			\STATE Solve the optimization problem~\eqref{eqn:fin}, determine the optimal $g_i^*$ and $u_i^*$.
			\STATE Inject the first group of system input in $u_i^*$ to the original open-loop system, and measure the system output $y_i$.
			\ENDFOR
		\end{algorithmic}
	\end{algorithm}
	
	\section{Illustrative Example}
	To clearly demonstrate the effectiveness of the proposed approach, an illustrative example of a multi-UAV system is used. In this example, the number of UAVs is set as $N=8$, and each UAV is controlled by 4 motors. 
	All the UAVs share the same dynamics, and the state-space model of each UAV is given by~\eqref{eq:dynamics},
	where $x=[
	p_x \,\,\, p_y \,\,\, p_z \,\,\, \dot p_x \,\,\, \dot p_y \,\,\,\dot p_z \,\,\, \omega_x \,\,\, \omega_y \,\,\,\omega_z \,\,\,\dot \omega_x \,\,\, \dot \omega_y \,\,\,\dot \omega_z ]^T$,
	$u=[
	u_1 \,\,\,u_2 \,\,\,u_3 \,\,\,u_4]
	^T 
	$.
	In the state vector, $p_x$, $p_y$, $p_z$ represent the spatial coordinates, $\omega_x$, $\omega_y$, $\omega_z$ represent the angular coordinates, $u_1$, $u_2$, $u_3$, $u_4$ denote the thrust of 4 motors. 
	It is assumed that all the state variables are measurable. The nominal values of matrices $A$, $B$ are obtained by linearizing the model of the UAV based on the Euler Method. Based on the model in~\cite{coulson2019regularized},
	matrices $A$, $B$ are given as follows:
	\begin{IEEEeqnarray}{l}
		A = \left[\begin{smallmatrix}
			1 & 0 & 0 & 0.1 & 0 & 0 & 0 & 0.049 & 0 & 0 & 0.0016 & 0 \\
			0 & 1 & 0 & 0 & 0.1 & 0 & -0.049 & 0 & 0 & -0.0016 & 0 & 0 \\
			0 & 0 & 1 & 0 & 0 & 0.1 & 0 & 0 & 0 & 0 & 0 & 0 \\
			0 & 0 & 0 & 1 & 0 & 0 & 0 & 0.981 & 0 & 0 & 0.049 & 0 \\
			0 & 0 & 0 & 0 & 1 & 0 & -0.981 & 0 & 0 & 0.049 & 0 & 0 \\
			0 & 0 & 0 & 0 & 0 & 1 & 0 & 0 & 0 & 0 & 0 & 0 \\
			0 & 0 & 0 & 0 & 0 & 0 & 1 & 0 & 0 & 0.1 & 0 & 0 \\
			0 & 0 & 0 & 0 & 0 & 0 & 0 & 1 & 0 & 0 & 0.1 & 0 \\
			0 & 0 & 0 & 0 & 0 & 0 & 0 & 0 & 1 & 0 & 0 & 0.1 \\
			0 & 0 & 0 & 0 & 0 & 0 & 0 & 0 & 0 & 1 & 0 & 0 \\
			0 & 0 & 0 & 0 & 0 & 0 & 0 & 0 & 0 & 0 & 1 & 0 \\
			0 & 0 & 0 & 0 & 0 & 0 & 0 & 0 & 0 & 0 & 0 & 1
		\end{smallmatrix}\right]\IEEEnonumber\\
		B =\left[\begin{smallmatrix}
			-2.3 \times 10^{-5} & 0 & 2.3 \times 10^{-5} & 0 \\
			0 & -2.3 \times 10^{-5} & 0 & 2.3 \times 10^{-5} \\
			1.75 \times 10^{-2} & 1.75 \times 10^{-2} & 1.75 \times 10^{-2} & 1.75 \times 10^{-2} \\
			-9.21 \times 10^{-4} & 0 & 9.21 \times 10^{-4} & 0 \\
			0 & -9.21 \times 10^{-4} & 0 & 9.21 \times 10^{-4} \\
			0.35 & 0.35 & 0.35 & 0.35 \\
			0 & 2.8 \times 10^{-3} & 0 & -2.8 \times 10^{-3} \\
			-2.8 \times 10^{-3} & 0 & 2.8 \times 10^{-3} & 0 \\
			3.7 \times 10^{-3} & -3.7 \times 10^{-3} & 3.7 \times 10^{-3} & -3.7 \times 10^{-3} \\
			0 & 5.6 \times 10^{-2} & 0 & -5.6 \times 10^{-2} \\
			-5.6 \times 10^{-2} & 0 & 5.6 \times 10^{-2} & 0 \\
			7.3 \times 10^{-2} & -7.3 \times 10^{-2} & 7.3 \times 10^{-2} & -7.3 \times 10^{-2}
		\end{smallmatrix}\right].\nonumber
	\end{IEEEeqnarray}
	Also, $C$ is an identity matrix, $D$ is a zero matrix. However, as mentioned above, the accuracy of the nominal model cannot be guaranteed in practical use. Here, the actual model under the perturbation is given by   
	\begin{IEEEeqnarray}{l}
		A_p = \left[\begin{smallmatrix}
			1 & 0 & 0 & 0.102 & 0 & 0 & 0 & 0.045 & 0 & 0 & 0.0015 & 0 \\
			0 & 1 & 0 & 0 & 0.097 & 0 & -0.05 & 0 & 0 & -0.0014 & 0 & 0 \\
			0 & 0 & 1 & 0 & 0 & 0.09 & 0 & 0 & 0 & 0 & 0 & 0 \\
			0 & 0 & 0 & 1 & 0 & 0 & 0 & 0.916 & 0 & 0 & 0.0505 & 0 \\
			0 & 0 & 0 & 0 & 1 & 0 & -1.03 & 0 & 0 & 0.0504 & 0 & 0 \\
			0 & 0 & 0 & 0 & 0 & 1 & 0 & 0 & 0 & 0 & 0 & 0 \\
			0 & 0 & 0 & 0 & 0 & 0 & 1 & 0 & 0 & 0.099 & 0 & 0 \\
			0 & 0 & 0 & 0 & 0 & 0 & 0 & 1 & 0 & 0 & 0.101 & 0 \\
			0 & 0 & 0 & 0 & 0 & 0 & 0 & 0 & 1 & 0 & 0 & 0.096 \\
			0 & 0 & 0 & 0 & 0 & 0 & 0 & 0 & 0 & 1 & 0 & 0 \\
			0 & 0 & 0 & 0 & 0 & 0 & 0 & 0 & 0 & 0 & 1 & 0 \\
			0 & 0 & 0 & 0 & 0 & 0 & 0 & 0 & 0 & 0 & 0 & 1
		\end{smallmatrix}\right].\IEEEnonumber\\\nonumber
	\end{IEEEeqnarray}
	It is pertinent to note that the matrix $A_p$ is only used to collect the closed-loop input-output data but not used in the calculation, because the actual value of $A_p$ is regarded as unknown in this work.

	\textcolor{black}{Since all the agents share the same dynamics, the following settings are the same for all agents, and thus the subscript $i$ will be omitted in the corresponding symbol. First, we set $T_{p}=1$, $T_{f}=30$. The input is persistently exciting of order $L=T_{p}+T_{f}+n=43$. Then, the minimum number of input/output measurements are $T_\textup{min}=(m+1)\times L-1=214$. Here, we choose $T_\textup{num}=214$. In this case, it ensures that the column span of the generalized Hankel matrix with respect to the input/output measurement data is the whole behavior space in the behavioral view. In our work, the sampling time is chosen as $0.1\,\textup{s}$.}
	
	\textcolor{black}{During the input/output measurement procedures, the same feedback controller is implemented to stabilize the closed-loop system for each agent. Notice that even though the precise system model is not available, we can still design a simple controller to stabilize an unstable system. In our work, we use an LQR controller for stabilization of the closed-loop system, such that the output $y_d$ of the resulting closed-loop system is bounded.} 
	Then, the matrix $W_i\in\mathbb{R}^{496\times 184}, \forall i =1,2,\cdots,N$ is effectively derived for each agent. Because the dimension of $W_i$ is quite large, the results will be not displayed in this paper. However, to show the effectiveness of closed-loop input/output measurements, a comparison is carried out by computing $\|W_i\|_2$ and $\|W_i\|_\infty$, where the detailed results are depicted in Table I. In this table, our proposed approach is denoted by Method 1, and the method with open-loop input/output measurements as presented in~\cite{coulson2019data} is denoted by Method 2. As can be seen, $\|W_i\|_2$ and $\|W_i\|_\infty$ obtained in our approach are apparently smaller, and it is certain because the system is stabilized in the input/output measurement procedure, and this phenomenon aligns well with our claims in Sec. III. 
	
	\begin{table}[t]
		\centering
		\caption{Comparison results of $\|W_i\|_2$ and $\|W_i\|_\infty$}
		\begin{tabular}{|c|c|c|c|c|}
			\hline
			\multirow{2}{*}{Agent} & \multicolumn{2}{c|}{$\|W_i\|_2$} & \multicolumn{2}{c|}{$\|W_i\|_\infty$} \\ \cline{2-5} 
			& Method 1     & Method 2     & Method 1        & Method 2        \\ \hline
			$i=1$                      & 110.22       & 61539.88     & 273.44          & 117792.93       \\ \hline
			$i=2$                      & 114.50       & 88612.26     & 282.80          & 137775.27       \\ \hline
			$i=3$                      & 107.82       & 49719.70     & 266.37          & 113771.13       \\ \hline
			$i=4$                      & 110.29       & 69574.57     & 271.79          & 112399.19       \\ \hline
			$i=5$                      & 109.25       & 60459.80     & 271.51          & 107693.96       \\ \hline
			$i=6$                      & 105.78       & 80080.19     & 263.99          & 114535.41       \\ \hline
			$i=7$                      & 111.22       & 60629.34     & 276.03          & 114074.74       \\ \hline
			$i=8$                      & 108.75       & 50473.13     & 270.28          & 113143.71       \\ \hline
		\end{tabular}
	\end{table}

\begin{figure}[t]
		\centering
		\includegraphics[trim=1cm 2cm 1cm 2cm, width=1\linewidth]{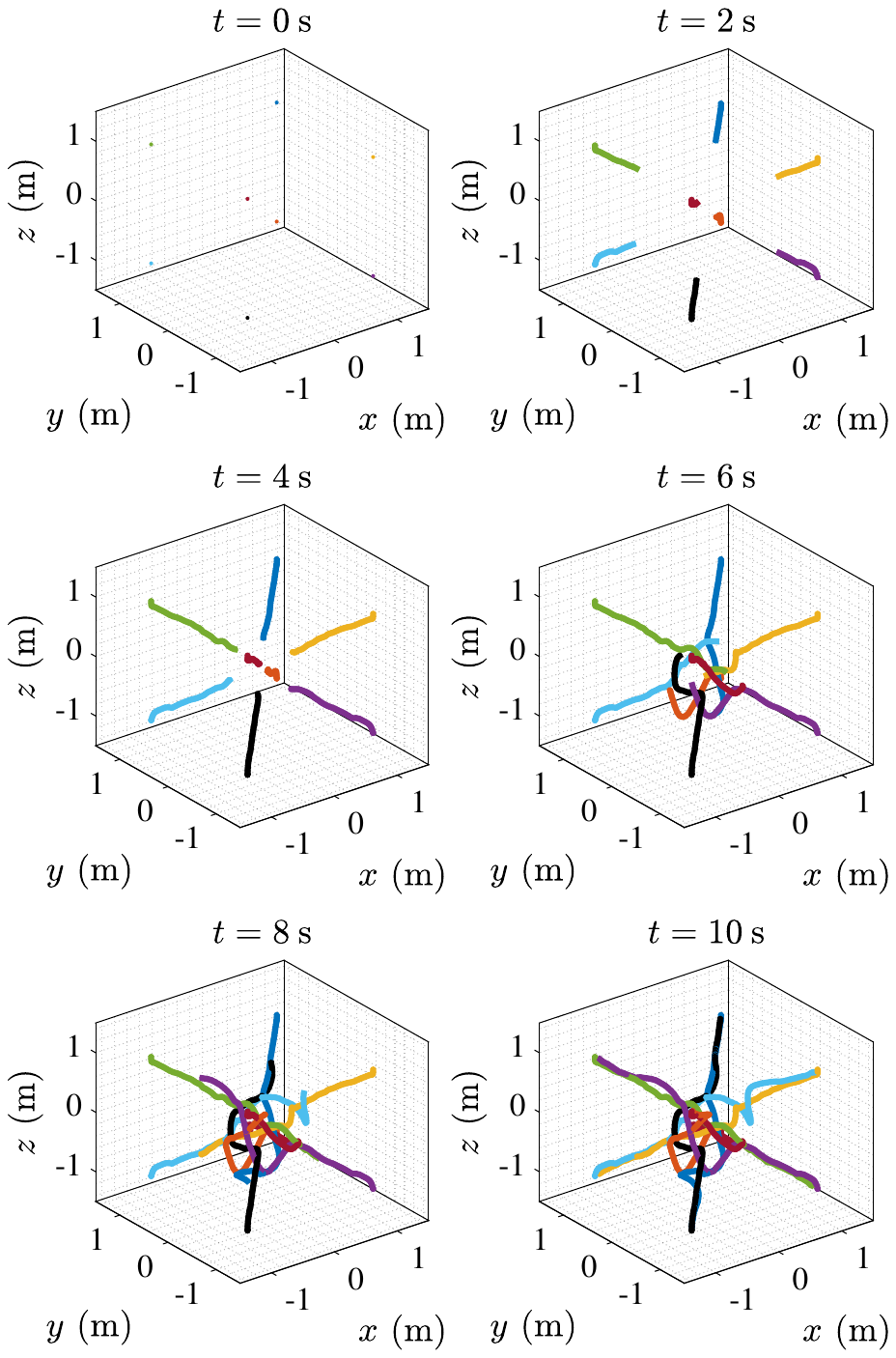}
		\caption{Resulted trajectories of all agents with the proposed approach.}
		\label{fig:Trajectory_3D}
	\end{figure}

	The parameters for the optimization problem~\eqref{eqn:fin} are given as follows. For all $i=1,2,\dotsm,N$, the weighting matrix $Q_i$ is chosen as $Q_i=I_T\otimes\operatorname{diag}(1,1,1,0,0,0,0,0,0,0,0,0)$, and $R_i$ is chosen as a zero matrix, which means only the tracking error is penalized in the objective function. $d_{\textup{safe}}$ is chosen as $0.5\,\textup{m}$. Moreover, there is no constraint imposed on the system output. The lower bound and upper bound of the control input are chosen as $-0.7007\,\textup{N}$ and $0.2993\,\textup{N}$, respectively. 
	For demonstrative purposes, the initial positions of 8 agents are located at different vertices of a cube, and each agent aims to move towards its corresponding diagonal vertex without making any collision. For example, the destination of the agent that initially located at $(-1\,\textup{m},-1\,\textup{m},-1\,\textup{m})$ is considered as $(1\,\textup{m},1\,\textup{m},1\,\textup{m})$. The simulation is implemented in a desktop platform with Intel(R) Xeon(R) W-2225 CPU @ 4.10GHz. The constrained data-driven predictive control optimization problem is solved by Casadi. By implementing the controller determined from the proposed approach, real-time trajectories of all the agents are illustrated in Fig.~\ref{fig:Trajectory_3D}. The 3D illustration clearly shows that the control objective in terms of path tracking is achieved, and no collision occurs in the whole process.
	
\begin{figure}[t]
		\centering
		\includegraphics[trim=1cm 2cm 1cm 2cm, width=1\linewidth]{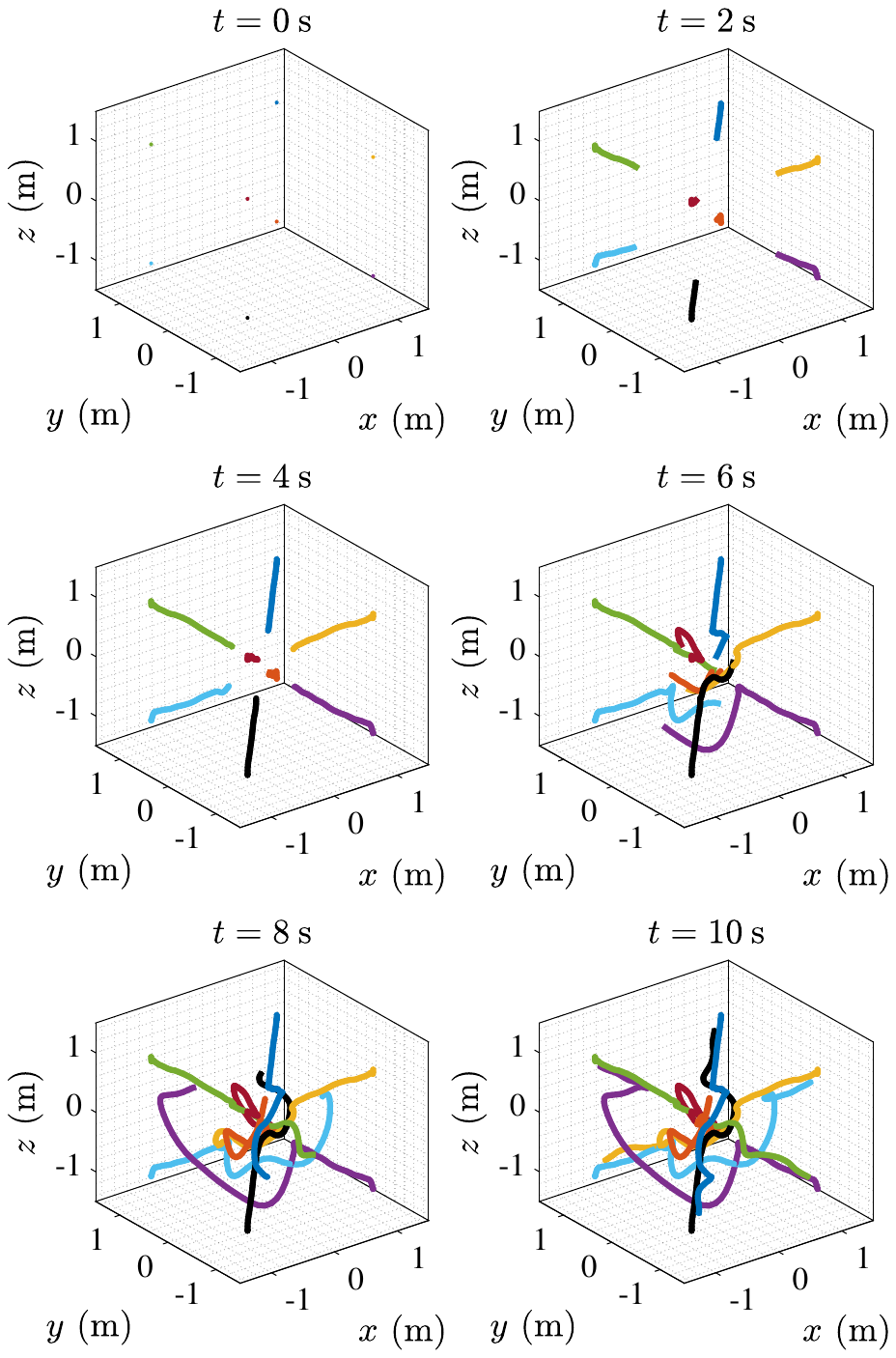}
		\caption{\textcolor{black}{Resulted trajectories of all agents with the conventional MPC.}}
		\label{fig:Trajectory_3D_MPC}
	\end{figure}

	In contrast, the conventional MPC requires the system model, which is generally obtained by system identification. Therefore, only nominal model can be used in the conventional MPC. In our work, additional comparisons are performed, and the same task for the multi-UAV system is conducted.
	The resulted trajectories of all the agents in 3D view are illustrated in Fig.~\ref{fig:Trajectory_3D_MPC}. With the conventional MPC, the control objective is also achieved without any collision. Compared with the results shown in Fig.~\ref{fig:Trajectory_3D} that are attained using our proposed approach, there are significant differences in resulted trajectories, and this is due to the different process in non-convex optimization involving the data or the model. To clearly quantify the difference in performance of the two approaches, the cost is determined. With the proposed approach, the cost is given by 48.9363; while with the conventional MPC, the cost is given by 99.1982. Compared with the conventional MPC, our proposed approach reduces the cost by 50.67\%. It is straightforward that our proposed data-driven approach outperforms the conventional model-based approach in this example. The main reason for this phenomenon is that our proposed method optimizes the predefined objective function based on the data rather than the nominal model. As a result, the optimum can still be obtained with our approach when the system model is not accurate. Therefore, compared to the conventional model-based MPC method, our proposed data-driven approach guarantees robustness against model uncertainties/perturbations. With the descriptions and pertinent analysis above, the effectiveness of the proposed approach is appropriately demonstrated.

	\section{Conclusion}
	In this work, a data-driven predictive control approach is presented to solve the multi-agent motion planning problem,
	where a non-parametric closed-loop behavior learning routine is presented for motion planning of multi-agent systems without explicit knowledge of the system model. In the approach here, a finite data set is collected offline under a closed-loop control framework, and an optimization problem is iteratively solved. With this framework, the optimal solution to the motion planning problem is efficiently and effectively determined. Finally, an example of multi-UAV system is introduced, and the simulation results clearly demonstrate the effectiveness of the proposed approach.
	
	\section*{Acknowledgement}
	
	The authors would like to thank Dr. Jeremy Coulson for his sharing and discussion.

\bibliographystyle{IEEEtran}
\bibliography{IEEEabrv,Reference}

\end{document}